\documentclass[10pt,conference]{IEEEtran}
\IEEEoverridecommandlockouts

\usepackage[colorlinks=false, hidelinks]{hyperref}
\usepackage{balance}
\usepackage{algorithm}
\usepackage[noend]{algpseudocode}

\usepackage{amsmath}
\usepackage{amsfonts}
\usepackage{enumitem}
\usepackage{graphicx}
\usepackage{subcaption}
\usepackage{listings}
\usepackage{xcolor}
\usepackage{colortbl}
\usepackage{pgfplots}
\usepackage{changepage}
\usepackage{multirow}
\usepackage{url}
\usepackage{tablefootnote}
\usepackage{threeparttable}
\usepackage[mathscr]{euscript}
\usepackage[normalem]{ulem}
\usepackage{xspace}

\AtBeginDocument{%
  } 

\definecolor{mygood}{RGB}{102, 158, 65}
\definecolor{mybad}{RGB}{242, 98, 105}

\definecolor{myblue}{RGB}{66, 135, 245}

\makeatletter
\newcommand*\myalgsize{%
  \@setfontsize\myalgsize{8}{9}%
}
\makeatother

\begin{document}

\title{FPPS: An FPGA-Based Point Cloud Processing System}

\author{
    \IEEEauthorblockN{Xiaofeng Zhou, Linfeng Du, Hanwei Fan, and Wei Zhang}
    \IEEEauthorblockA{
        Department of ECE, Hong Kong University of Science and Technology\\
        \{xzhoubu, linfeng.du, hfanah\}@connect.ust.hk, wei.zhang@ust.hk
    }
}
\maketitle
\begin{abstract}


Point cloud processing is a computational bottleneck in autonomous driving systems, especially for real-time applications, while energy efficiency remains a critical system constraint. This work presents FPPS, an FPGA-accelerated point cloud processing system designed to optimize the iterative closest point (ICP) algorithm, a classic cornerstone of 3D localization and perception pipelines. Evaluated on the widely used KITTI benchmark dataset, the proposed system achieves up to 35$\times$ (and a runtime-weighted average of 15.95$\times$) speedup over a state-of-the-art CPU baseline while maintaining equivalent registration accuracy. Notably, the design improves average power efficiency by 8.58$\times$, offering a compelling balance between performance and energy consumption. These results position FPPS as a viable solution for resource-constrained embedded autonomous platforms where both latency and power are key design priorities.

\end{abstract} 
\begin{IEEEkeywords}
FPGA acceleration, point cloud registration, low-power computing, real-time perception.
\end{IEEEkeywords}
\section{Introduction}\label{sec:intro}
Point cloud registration has become a foundational component in autonomous driving systems, enabling critical tasks such as simultaneous localization and mapping, object detection, and motion planning. Among the various algorithms employed, the \textit{iterative closest point (ICP)}~\cite{ICP1992} algorithm remains a widely adopted method for geometric alignment due to its effectiveness. However, the computational demands of ICP pose significant challenges for embedded platforms constrained by power, size, and thermal budgets, especially in real-time scenarios that require processing high-resolution point clouds with low latency.

Autonomous driving demands point cloud registration with strictly low-latency responsiveness at sustained high throughput. This latency sensitivity stems from the safety-critical nature of real-time environment perception---where dynamic obstacles and unpredictable scenarios require sub-frame reaction times for collision avoidance and path planning. While CPU implementations achieve throughput scaling through massive multi-core parallelism, their non-linear power increase and memory contention effects fundamentally compromise latency determinism, creating unacceptable risks for autonomous decision-making. 
To address these limitations, this work proposes an FPGA-based point cloud registration system that integrates a \textit{field-programmable gate array (FPGA)} with a host CPU to offload and parallelize the most computationally intensive stages of the ICP pipeline. 
Compared to ASIC implementations, FPGAs offer greater flexibility for evolving algorithm designs in autonomous driving, significantly accelerating development cycles and reducing time-to-market.
To facilitate integration and usability, we additionally developed a set of PCL-like~\cite{PCL2011} APIs that abstract the underlying hardware operations. These APIs provide a familiar programming model for developers, enabling rapid prototyping and seamless adoption within existing PCL-based point cloud registration workflows.

The remainder of this design summary is organized as follows: Section~\ref{sec:pre} introduces the ICP algorithm; Section~\ref{sec:method} describes the FPGA-based system design in detail; Section~\ref{sec:res} demonstrates the performance and resource consumption results of our design on the target FPGA; Section~\ref{sec:related} describes the related works. Section~\ref{sec:discussion} discusses potential improvements for our design, and Section~\ref{sec:conclusion} concludes the paper.


\section{Preliminaries}\label{sec:pre}

\subsection{Point cloud registration}

\begin{figure}[t]
    \centering
    \includegraphics[width=0.95\linewidth]{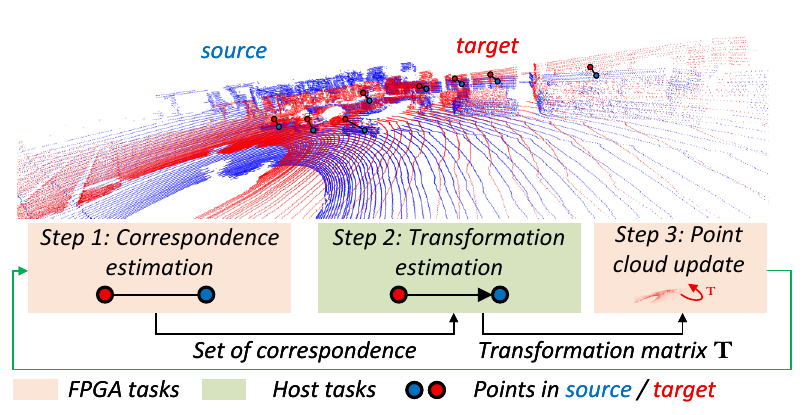}
    \caption{An overview of point cloud registration.}
    \label{fig:icp}
\end{figure}

Point cloud registration tasks aim at aligning two sets of 3D points into a common coordinate frame, as shown in Figure~\ref{fig:icp}. The point sets are typically referred to as the \textit{source} and \textit{target} point clouds, represented in the blue and red points, respectively, in Figure~\ref{fig:icp}.  Among various registration techniques, the ICP algorithm remains one of the most widely used due to its simplicity and effectiveness. ICP operates by iteratively refining a rigid transformation (calculated by step 2 in Figure~\ref{fig:icp}) that minimizes the distance between corresponding points in the source and target clouds. A rigid transformation is a geometric operation that preserves the shape and size of an object by applying only rotation, translation, or reflection, without any scaling or distortion.

Formally, given a \textit{source} point cloud \( \mathcal{P} = \{ \mathbf{p}_i \in \mathbb{R}^3 \} \) and a \textit{target} point cloud \( \mathcal{Q} = \{ \mathbf{q}_i \in \mathbb{R}^3 \} \), the objective is to find the optimal rotation matrix \( \mathbf{R} \in SO(3) \) and translation vector \( \mathbf{t} \in \mathbb{R}^3 \) that minimize the following cost function. Here, $SO(3)$ represents all possible rotations in 3D space that preserve orientation and distance.  Mathematically, it consists of all 3$\times$3 orthogonal matrices with determinant equal to 1.

\begin{equation}
E(\mathbf{R}, \mathbf{t}) = \sum_{i=1}^{N} \left\| \mathbf{q}_i - (\mathbf{R} \mathbf{p}_i + \mathbf{t}) \right\|^2
\end{equation}

where \( \mathbf{p}_i \) and \( \mathbf{q}_i \) are corresponding points in the source and target clouds, respectively, and \( N \) is the number of point correspondences. A pair of corresponding points is a pair of points, each from $\mathcal{P}$ and $\mathcal{Q}$ that represents the same physical location in a scene. The transformation matrix derived in each iteration $\mathbf{T}_j$ is formed by augmenting rotation $\mathbf{R}_j$ and translation $\mathbf{t}_j$, where $j$ is the iteration number:

\begin{equation}
    \mathbf{T}_j = \begin{bmatrix}
\mathbf{R}_j & \mathbf{t}_j \\
\mathbf{0} & 1
\end{bmatrix}
\end{equation}

The ICP algorithm typically proceeds in the following steps:

\begin{enumerate}
    \item \textbf{Correspondence Estimation:} For each point \( \mathbf{p}_i \in \mathcal{P} \), find the closest point \( \mathbf{q}_i \in \mathcal{Q} \) using a \textit{nearest-neighbor (NN)} search.
    \item \textbf{Transformation Estimation:} Compute the optimal rigid transformation \( (\mathbf{R}, \mathbf{t}) \) that minimizes the cost function that represents alignment error using \textit{singular value decomposition (SVD)}.
    \item \textbf{Point cloud update:} Update the source point cloud by applying the estimated transformation.
    \item \textbf{Convergence Check:} Repeat the process until convergence criteria are met, such as a maximum number of iterations or a threshold on the transformation matrix: When $\mathbf{R}$ is close to the identity matrix and $\mathbf{t}$ is close to the zero vector, it can be ignored as we have found a rigid transformation from $\mathcal{P}$ to $\mathcal{Q}$.
\end{enumerate}

The final optimal transformation $\mathbf{T}$ is derived by cumulatively applying the transformation matrix derived in each iteration:

\begin{equation}
    \mathbf{T} = \prod_{j} \mathbf{T}_j
\end{equation}

Despite its effectiveness, ICP is computationally intensive due to the repeated NN searches and matrix operations. This makes deployment challenging on embedded platforms with limited computational resources and strict power budgets. Therefore, FPGAs offer a promising solution to achieve real-time performance while maintaining energy efficiency.


\section{Method}\label{sec:method}

This section describes the hardware design of the proposed point cloud processing system. The architecture is optimized for low-latency, power-efficient point cloud registration through pipelined transformation and parallel NN search.

\subsection{Hardware Architecture}

\begin{figure}[htbp]
    \centering
    \includegraphics[width=0.8\linewidth]{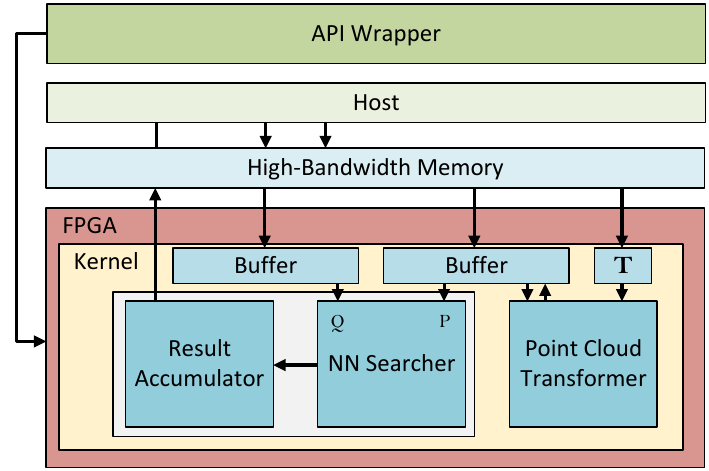}
    \vspace{-0.5\baselineskip}
    \caption{The system diagram of FPPS.}
    \label{fig:hw_block_diagram}
\end{figure}

Figure \ref{fig:hw_block_diagram} demonstrates the system diagram of FPPS. It includes a CPU host that is responsible for data transmission and invokes kernel execution according to the instructions from APIs. A point cloud processing kernel is implemented on the FPGA part of the SoC, which includes a point cloud transformer according to $\mathbf{T}$, an NN searcher to search the NN in $\mathcal{Q}$ for each $p_i$ in a point cloud, and a result accumulator to calculate the covariance matrix for \textit{singular value decomposition (SVD)} according to the NN result.

Two on-chip buffers are used to store the point cloud data for NN and point cloud transformation. To minimize the memory access latency, during the point cloud processing, all data in the point cloud is stored in on-chip memory. The host and the kernel communicate through a high-bandwidth memory interface. 

\subsection{NN Searcher}

\begin{figure}[htbp]
    \centering
    \includegraphics[width=0.95\linewidth]{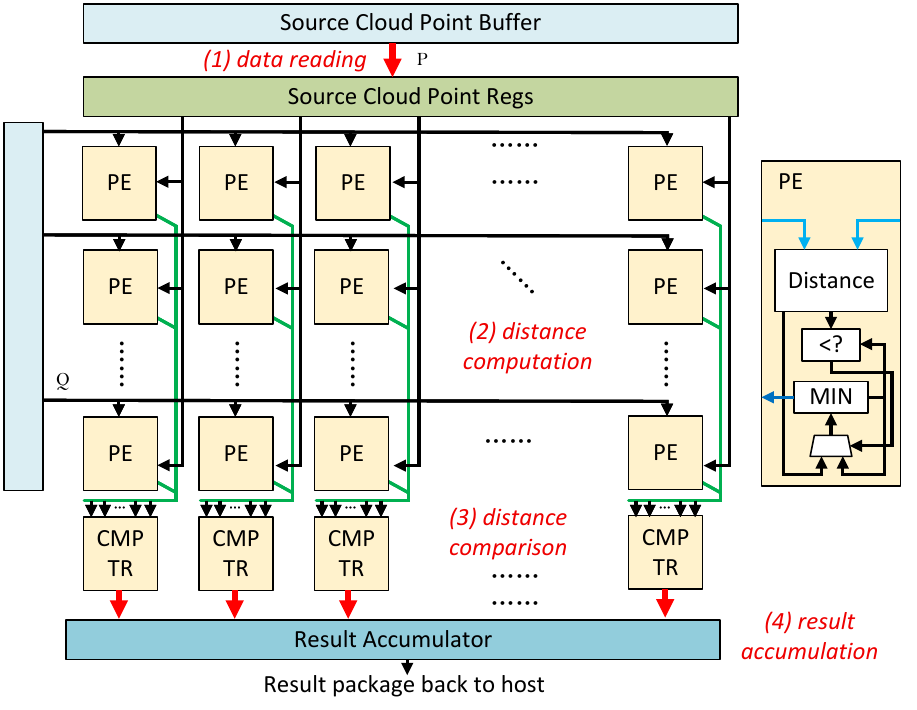}
    \vspace{-0.5\baselineskip}
    \caption{The task-level pipelined design of NN searcher.}
    \label{fig:detailed_design}
\end{figure}

The core structure inside the point cloud processing kernel is the NN searcher, which implements the most computationally intensive part of the ICP algorithm. The design diagram of the NN searcher is shown in Figure~\ref{fig:detailed_design}.
The architecture adopts a streaming model, where data flows through pipelined stages connected by FIFO buffers. The streaming model is divided into four stages that execute concurrently: (1)~data reading, (2)~distance computation, (3)~distance comparison, and (4)~result accumulation. The data reading part reads the cloud point data from the global kernel buffer implemented in BRAM to the local register buffer. Once the local source cloud point register buffer is full, the data will pass to the distance computation stage. In this stage, the downstream processing array will calculate the distance between each pair of the source cloud point and the destination cloud point in parallel. Different from the source cloud points, which are collected by a local register buffer to enable parallel execution, the storage of destination cloud points is partitioned into pieces, such that a batch of points can be read and broadcast to the distance computation array in parallel. The computed distances are stored locally in the \textit{processing element (PE)} array, whose implementation is shown on the right part of the Figure~\ref{fig:detailed_design}. The \texttt{Distance} block calculates the distance between two points. The \texttt{MIN} block consists of two registers that store the current minimum distance value and the NN candidate, respectively. Once a destination point with a smaller distance arrives, the \texttt{MIN} block will be updated. In our target scenario, we can process around 130k NN candidates for each cloud point. 

The distance comparison stage makes a competition among the NN candidates for each column of PEs. A group comparison tree (CMP TR) is implemented to derive the final NN for each source point. Finally, in the result accumulation stage, the source cloud point will be sent to the result accumulator together with its NN in the destination point cloud, also in a streaming manner.

\begin{table*}[htbp]
\centering
\begin{tabular}{|p{5.5cm}|p{7cm}|p{4.2cm}|}
\hline
\textbf{Function Name} & \textbf{Description} & \textbf{Arguments} \\
\hline
\texttt{hardwareInitialize()} & Initializes the hardware and loads the \texttt{.xclbin}.  & None \\
\hline
\texttt{setTransformationMatrix()} & Sets the initial transformation matrix. The matrix is fed into the FPGA and used for one transformation before ICP. & \texttt{transformationMatrix} \\
\hline
\texttt{setInputSource()} & Sets the source point cloud for the ICP algorithm. & \texttt{inputSource} \\
\hline
\texttt{setInputTarget()} & Sets the target point cloud for the ICP algorithm. & \texttt{inputTarget} \\
\hline
\texttt{setMaxCorrespondenceDistance()} & Sets the maximum correspondence distance to filter out outliers in the ICP algorithm. & \texttt{maxCorrespondenceDistance} \\
\hline
\texttt{setMaxIterationCount()} & Sets the maximum number of iterations for the ICP algorithm. & \texttt{maxIterationCount} \\
\hline
\texttt{setTransformationEpsilon()} & Sets the stopping criteria for the ICP algorithm iteration using the epsilon of the transformation matrix. & \texttt{transformationEpsilon} \\
\hline
\texttt{align()} & Performs the alignment procedure and returns the final transformation matrix. & None \\
\hline
\end{tabular}
\caption{API Summary of FPPS with Arguments}
\vspace{-0.5\baselineskip}
\label{tab:xf_icp_api_args}
\end{table*}

\section{Evaluation}\label{sec:res}

\subsection{Experimental Setup}

To evaluate the performance of the proposed FPGA-accelerated ICP framework, we conduct experiments using a set of workloads derived from the KITTI dataset~\cite{KITTI2013IJRR}.  The hardware prototype is implemented on an AMD Alveo U50 accelerator card, linked to a host CPU for hybrid execution. The host system is equipped with an 11th Gen Intel Core i5-11400 processor running at 2.60\,GHz. For baseline comparison, a high-performance Intel Xeon Gold 6246R CPU at 3.40\,GHz is used to execute a software-only ICP implementation based on PCL~\cite{PCL2011}.  We use PowerTOP v2.11 to measure the host CPU power.

We use the KITTI odometry dataset to benchmark the system under realistic autonomous driving scenarios. For each frame, 4096 points are randomly sampled from the source point cloud to initialize the ICP alignment. The full point cloud is then processed through global ICP iterations to compute the final transformation.

The ICP configuration is fixed across all experiments with the following parameters:
\begin{itemize}
    \item Maximum number of iterations: 50
    \item Maximum correspondence distance: 1.0\,m
    \item Transformation convergence threshold: $1 \times 10^{-5}$
\end{itemize}

Performance is measured in terms of average \textit{root mean square error (RMSE)}, per-frame latency, and computational power consumption.

\subsection{Resource Utilization}

\begin{figure}
    \centering
    \includegraphics[width=0.9\linewidth]{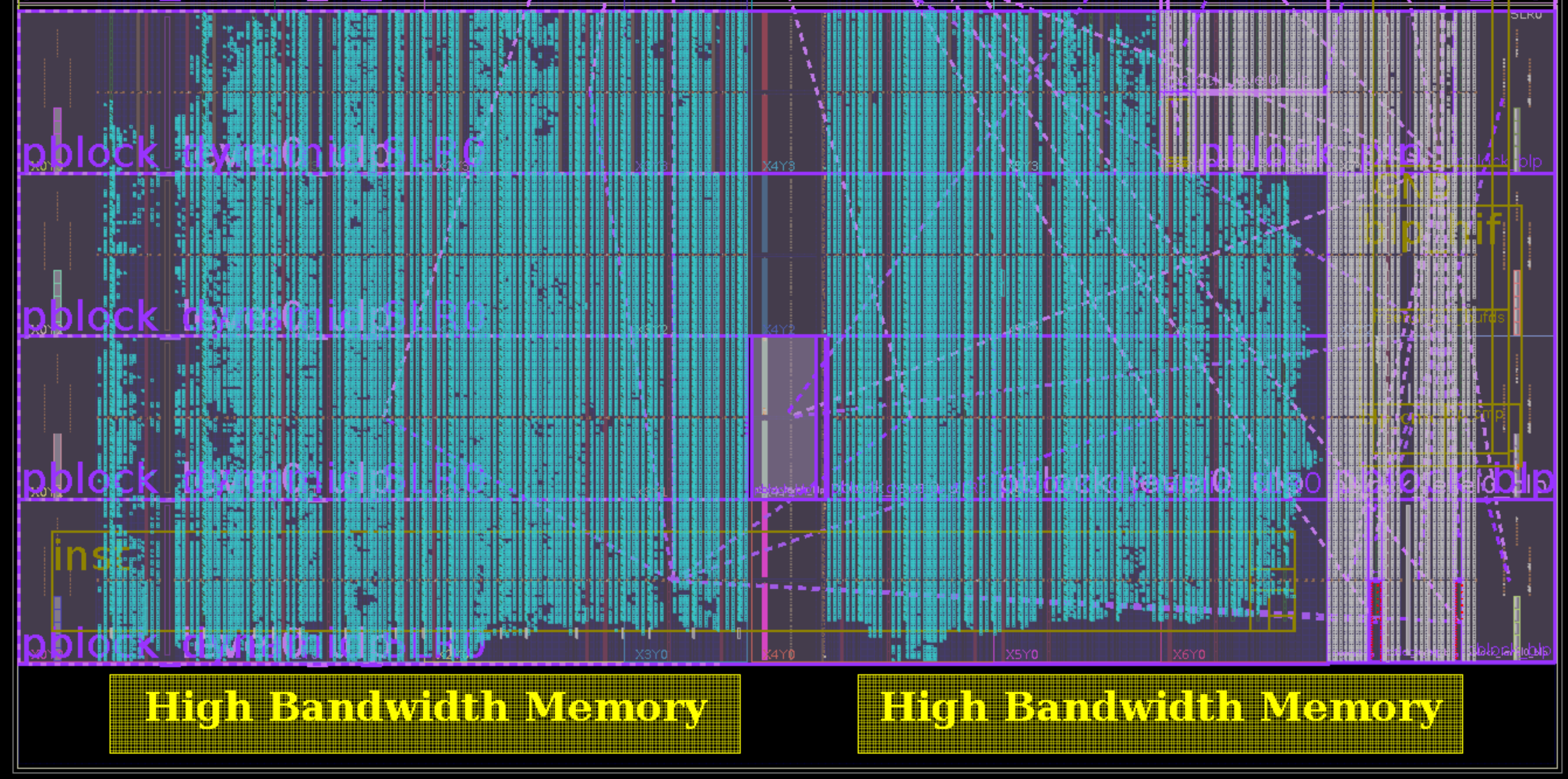}
    \caption{Post-routing device view of the accelerator on Alveo U50 (\texttt{SLR0}).}
    \label{fig:slr0}
\end{figure}

Table~\ref{tab:resource} summarizes the hardware resource utilization of the proposed FPGA-based point cloud processing accelerator, with the corresponding post-routing device view of \texttt{SLR0} shown in Figure~\ref{fig:slr0}.  The design occupies one of the two \textit{super logic regions (SLRs)} on U50 for the sake of communication efficiency with hosts. On U50 accelerator card, only \texttt{SLR0} can access the HBM that connects to the host. Resource breakdown is shown in Table~\ref{tab:resource}.

\begin{table}[ht]
\centering
\caption{FPGA resource usage summary}
\begin{tabular}{|c|c|c|c|}
\hline
\textbf{Resource} & \textbf{Usage}  & \textbf{Utilization on \texttt{SLR0}} & \textbf{Overall Utilization}\\
\hline
LUT & 313,542 & 71.94\% & 36.04\%  \\
\hline
FF & 441,273 & 50.62\% & 25.36\%   \\
\hline
Block RAM & 613 & 45.61\% & 22.80\% \\
\hline
DSP & 2,384 &  80.11\%  & 40.13\% \\
\hline
\end{tabular}
\vspace{-1.5\baselineskip}
\label{tab:resource}
\end{table}

\subsection{Performance and accuracy}

Table \ref{tab:rmse} presents the average RMSE across ten sequences from the KITTI dataset, comparing the baseline CPU implementation in PCL with the proposed system. The results indicate that the integration of FPGA acceleration in our design does not compromise registration accuracy with marginal variations within 0.01 meters. This consistency confirms that the proposed hardware design replicates the numerical behavior of the original software implementation, thereby validating its suitability for high-precision SLAM and perception tasks in autonomous driving.

\begin{table}[ht]
\centering
\caption{Average RMSE comparison (meter)}
\begin{tabular}{|c|ccccc|}
\hline
\textbf{Sequence} & 00 & 01 & 02 & 03 & 04 \\
\hline
\textbf{CPU} & 0.198 & 0.417 & 0.205 & 0.218 & 0.330 \\
\textbf{CPU+FPGA} & 0.265 & 0.422 & 0.205 & 0.218 & 0.329 \\
\hline
\end{tabular}

\vspace{0.1cm}

\begin{tabular}{|c|ccccc|}
\hline
\textbf{Sequence} & 05 & 06 & 07 & 08 & 09 \\
\hline
\textbf{CPU} & 0.197 & 0.271 & 0.178 & 0.216 & 0.215 \\
\textbf{CPU+FPGA} & 0.198 & 0.272 & 0.178 & 0.216 & 0.216 \\
\hline
\end{tabular}
\label{tab:rmse}
\end{table}

Table \ref{tab:lat} reports the average per-frame latency for baseline CPU implementation (in rows \textbf{CPU}) and the proposed implementation with FPGA (in rows \textbf{CPU+FPGA}). The hybrid system achieves substantial speedups, ranging from 4.82$\times$ to 35.36$\times$ across different sequences. These results underscore the effectiveness of the proposed architecture in reducing computational latency, a critical requirement for real-time operation in autonomous vehicles. The consistent acceleration across diverse scenarios highlights the robustness and generalizability of the design.

\begin{table}[ht]
\centering
\caption{Average latency per frame and acceleration rate}
\vspace{-0.5\baselineskip}
\begin{tabular}{|c|ccccc|}
\hline
\textbf{Sequence} & 00 & 01 & 02 & 03 & 04 \\
\hline
\textbf{CPU (ms)} & 3714.5 & 8640.1 & 1363.3 & 4820.2 & 2591.9 \\
\textbf{CPU+FPGA (ms)} & 162.6 & 537.4 & 237.2 & 136.3 & 537.4 \\
\textbf{Acceleration} & 22.84$\times$ & 16.07$\times$ & 5.75$\times$ & 35.36$\times$ & 4.82$\times$ \\
\hline
\end{tabular}

\vspace{0.1cm}
\begin{tabular}{|c|ccccc|}
\hline
\textbf{Sequence} & 05 & 06 & 07 & 08 & 09 \\
\hline
\textbf{CPU (ms)} & 3523.8 & 5213.9 & 3164.1 & 3662.7 & 7037.1 \\
\textbf{CPU+FPGA (ms)} & 148.7 & 224.3 & 145.1 & 136.3 & 477.6 \\
\textbf{Acceleration} & 23.69$\times$ & 23.24$\times$ & 21.80$\times$ & 26.87$\times$ & 14.7$\times$ \\
\hline
\end{tabular}
\label{tab:lat}
\end{table}

\subsection{Power efficiency}


The proposed FPGA design achieves 8.58$\times$ higher power efficiency (calculated by the ratio of power consumption against the execution speed) compared to the CPU baseline (16.3W), despite its 28W total power consumption (14W static + 14W dynamic power for FPGA, plus 2.3W host power). This demonstrates that the significant performance gains outweigh the moderate power increase, making the solution ideal for energy-constrained autonomous systems where performance-per-watt is paramount.

\section{Discussion}\label{sec:discussion}

\subsection{On the choice of NN search strategy}

While k-d trees are widely used for efficient NN search in point cloud registration~\cite{icpkd}, they were not adopted in our hardware design due to their sequential execution manner and high dependency across the tree search process for a single point. Preliminary experiments revealed that k-d tree traversal introduces significant latency, with average per-frame delays exceeding 250\,ms in some sequences. This is primarily due to the inherently sequential nature of tree traversal, which limits opportunities for parallelism and pipelining on FPGA architectures.

Approximate k-d tree search can reduce computational complexity but often leads to degraded convergence in ICP due to inaccurate correspondences. Conversely, exact search requires backward tracing, which further increases latency and complicates control logic. These trade-offs make k-d trees suboptimal for low-latency, deterministic hardware pipelines.

Instead, we adopt a fully parallel nearest neighbor search strategy based on a systolic array structure. This approach enables high-throughput, deterministic latency, and better compatibility with pipelined execution, making it more suitable for real-time deployment in embedded autonomous systems.

Although previous work on accelerating the ICP algorithm has been conducted under various scenarios, in this work, we aim to make the implementation of ICP the same in the PCL library in terms of functionality. The reason for this design choice is that we want to make it easier for software developers to use the system, as they can rely on familiar interfaces and functionalities.

\section{Related Works}\label{sec:related}

Recent FPGA-based ICP accelerators have targeted specific application scenarios. Kosuge et al.~\cite{kosuge20194} proposed a reused hierarchical graph and sorting-network-based generator for object picking robots, achieving 4.8× speedup but relying on static graph structures and limited adaptability. Their later work~\cite{2021kos} introduced partial reconfiguration to reduce resource usage, yet remains tailored to low-resolution, indoor scenes. Wang et al.~\cite{iscas24-knn} focused on mobile platforms, using locality sensitive hashing for fast KNN search, trading accuracy for speed. Our work differs by targeting latency-and precision-critical autonomous driving, accelerating the full ICP pipeline with exact, parallel NN search and pipelined transformation estimation. The design maintains PCL compatibility, enabling seamless integration into real-time perception systems.




\section{Conclusion}\label{sec:conclusion}

This design summary presents a hardware-accelerated framework for point cloud registration that leverages FPGA-based computation to address the latency and power limitations of traditional implementations. By offloading key stages of the ICP algorithm to an FPGA, the proposed design enables real-time performance while significantly reducing energy consumption. The results demonstrate the improvement of our FPGA-based design for 3D perception in autonomous systems.




\bibliographystyle{IEEEtran}
\balance
\bibliography{ICP/ICP.bib}

@String{Computing = "Computing" }

@article{ICP1992,
  author={Besl, P.J. and McKay, Neil D.},
  journal={IEEE Transactions on Pattern Analysis and Machine Intelligence}, 
  title={A method for registration of 3-D shapes}, 
  year={1992},
  volume={14},
  number={2},
  pages={239-256},
  doi={10.1109/34.121791}}

@INPROCEEDINGS{PCL2011,
  author={Rusu, Radu Bogdan and Cousins, Steve},
  booktitle={2011 IEEE International Conference on Robotics and Automation}, 
  title={3D is here: Point Cloud Library (PCL)}, 
  year={2011},
  volume={},
  number={},
  pages={1-4},
  doi={10.1109/ICRA.2011.5980567}}

@article{KITTI2013IJRR,
  author = {Andreas Geiger and Philip Lenz and Christoph Stiller and Raquel Urtasun},
  title = {Vision meets Robotics: The KITTI Dataset},
  journal = {International Journal of Robotics Research (IJRR)},
  year = {2013}
}

@INPROCEEDINGS{icpkd,
  author={Greenspan, M. and Yurick, M.},
  booktitle={Fourth International Conference on 3-D Digital Imaging and Modeling, 2003. 3DIM 2003. Proceedings.}, 
  title={Approximate k-d tree search for efficient ICP}, 
  year={2003},
  volume={},
  number={},
  pages={442-448},
  doi={10.1109/IM.2003.1240280}}

@inproceedings{kosuge20194,
  title={A 4.8 x faster FPGA-based iterative closest point accelerator for object pose estimation of picking robot applications},
  author={Kosuge, Atsutake and Yamamoto, Keisuke and Akamine, Yukinori and Yamawaki, Taizo and Oshima, Takashi},
  booktitle={2019 IEEE 27th Annual International Symposium on Field-Programmable Custom Computing Machines (FCCM)},
  pages={331--331},
  year={2019},
  organization={IEEE}
}

@INPROCEEDINGS{iscas24-knn,
  author={Wang, Chengliang and Huang, Zhetong and Ren, Ao and Zhang, Xun},
  booktitle={2024 IEEE International Symposium on Circuits and Systems (ISCAS)}, 
  title={An FPGA-based kNN Seach Accelerator for point cloud registration}, 
  year={2024},
  volume={},
  number={},
  pages={1-5},
  doi={10.1109/ISCAS58744.2024.10558303}}

@ARTICLE{2021kos,
  author={Kosuge, Atsutake and Yamamoto, Keisuke and Akamine, Yukinori and Oshima, Takashi},
  journal={IEEE Transactions on Industrial Electronics}, 
  title={An SoC-FPGA-Based Iterative-Closest-Point Accelerator Enabling Faster Picking Robots}, 
  year={2021},
  volume={68},
  number={4},
  pages={3567-3576},
  doi={10.1109/TIE.2020.2978722}}


\end{document}